\documentclass[aps,pre,showpacs,notitlepage,longbibliography]{revtex4-1}
\usepackage{epsfig,graphics,amssymb,amsmath,subeqnarray,color,bm,bbm, mathtools}
\usepackage[toc,page]{appendix}
\usepackage{mathtools}
\usepackage{kpfonts}
\usepackage{xcolor}
\usepackage{float}
 \usepackage[utf8]{inputenc}
\usepackage{graphicx}
\usepackage[colorlinks=true,linkcolor=blue,]{hyperref}
\usepackage[sfdefault=cmbr]{isomath}

\usepackage{scalerel,stackengine}
\stackMath
\newcommand\reallywidehat[1]{%
\savestack{\tmpbox}{\stretchto{%
  \scaleto{%
    \scalerel*[\widthof{\ensuremath{#1}}]{\kern-.6pt\bigwedge\kern-.6pt}%
    {\rule[-\textheight/2]{1ex}{\textheight}}
  }{\textheight}%
}{0.5ex}}%
\stackon[1pt]{#1}{\tmpbox}%
}
\usepackage{bm}
\def\boldsymbol{\bm}

\def \t{\tensorsym}
\def \lb{\left}
\def \rb{\right}

\def \d{\,\text{d}}

\def \De{\text{De}}

\def \Re{\text{Re}}

\def \etah{\hat{\eta}}

\def \bgamma{\boldsymbol{\gamma}}

\def \bgammad{\dot{\boldsymbol{\gamma}}}
\def \bgammadh{\hat{{\dot{\bgamma}}}}

\def \bnabla{\boldsymbol{\nabla}}

\def \bOmega{\bm{\Omega}}

\def \para{\parallel}

\def \bsigmah{\hat{\boldsymbol{\sigma}}}

\def \btau{\boldsymbol{\tau}}

\def \bzero{\bm{0}}

\def \bA{\bm{A}}

\def \fB{\mathcal{B}}

\def \bEh{\hat{\bm{E}}}

\def \tEh{\mathsf{\t{\hat{E}}}}

\def \bF{\bm{F}}

\def \tF{\mathsf{\t F}}
\def \tFh{\mathsf{\t{\hat{F}}}}

\def \bL{\bm{L}}

\def \bN{\bm{N}}

\def \fO{\mathcal{O}}

\def \Rh{\hat{R}}
\def \bR{\bm{R}}
\def \bRh{\hat{\bm{R}}}

\def \tR{\mathsf{\t R}}
\def \tRh{\mathsf{\t{\hat{R}}}}

\def \tTh{\mathsf{\t{\hat{T}}}}

\def \bU{\bm{U}}

\def \tU{\mathsf{\t U}}
\def \tUh{\mathsf{\t{\hat{U}}}}

\def \fV{\mathcal{V}}

\def \be{\bm{e}}

\def \bn{\bm{n}}

\def \br{\bm{r}}

\def \bv{\bm{v}}

\def \bx{\bm{x}}

\definecolor{navyblue}{RGB}{0,0,128}
\definecolor{dodgerblue}{RGB}{30,144,255}
\definecolor{darkgrey}{RGB}{169,169,169}
\definecolor{deepskyblue}{RGB}{0, 191, 255}

\usepackage{bm}
\def\boldsymbol{\bm}

\usepackage{graphicx}
\definecolor{navyblue}{RGB}{0,0,128}
\definecolor{dodgerblue}{RGB}{30,144,255}
\definecolor{darkgrey}{RGB}{169,169,169}
\definecolor{deepskyblue}{RGB}{0, 191, 255}
\newcommand{\di}{i}  
 \def\boldsymbol{\bm}

\begin{document}
\title{Two-sphere swimmers in viscoelastic fluids}
\author{Charu Datt, Babak Nasouri}
\affiliation{Department of Mechanical Engineering, Institute of Applied Mathematics 
University of British Columbia,
Vancouver, BC, V6T 1Z4, Canada
}

\author{Gwynn J. Elfring}\email{Electronic mail: gelfring@mech.ubc.ca}
\affiliation{Department of Mechanical Engineering, Institute of Applied Mathematics 
University of British Columbia,
Vancouver, BC, V6T 1Z4, Canada
}
\date{\today}

\begin{abstract}
We examine swimmers comprising of two rigid spheres which oscillate periodically along their axis of symmetry, considering both when the oscillation is in phase and anti-phase, and study the effects of fluid viscoelasticity on their net motion. These swimmers both display reciprocal motion in a Newtonian fluid and hence no net swimming is achieved over one cycle. Conversely, we find that when the two spheres are of different sizes, the effect of viscoelasticity acts to propel the swimmers forward in the direction of the smaller sphere. Finally, we compare the motion of rigid spheres oscillating in viscoelastic fluids with elastic spheres in Newtonian fluids where we find similar results.
\end{abstract}

\maketitle

\section{Introduction}
Recent review articles on swimming at small length scales \citep{Bechinger_review, eric_review, elgeti_review, active_brownian_review, Microrobot_review, Stocker_review} point to the immense interest in recent years on understanding the topic that has wide ranging applications from  biomedical engineering \cite{nano} to autonomous de-pollution of water and soil \cite{gao_2014}. Several theoretical models for understanding swimming at low Reynolds number in Newtonian fluids have been developed such as the swimming sheet \cite{taylor1951}, and the squirmer \cite{Lighthill1951}. The swimming techniques used in these two models, which were drawn from observing biological swimmers, demonstrate effective ways to circumvent the scallop theorem, which stipulates that a reciprocal swimming gait cannot lead to net motion at low Reynolds numbers in Newtonian fluids \cite{purcell1977}. Beyond the swimming sheet and the squirmer, other theoretical models have been proposed; many aiming simplicity. Purcell in his famous 1976 talk ``Life at low Reynolds number" proposed the ``simplest animal" that could swim: a planar three-linked swimmer, which could move by alternately moving its front and rear segments \cite{purcell1977, stone_purcell}.  The Najafi-Golestanian swimmer \cite{najafi2004} propels forward using its collinear assembly of three equal spheres, connected with thin rods which vary in lengths as the spheres oscillate in a non time-reversible way \cite{felderhof_animacules, golestanian2008, alexander2009}. \citet{avron2005} proposed another model, more efficient than the three-sphere model, where the swimmer consists of just a pair of spherical bladders which exchange their volumes while also varying their distance of separation. These models have been instrumental in understanding swimming at low Reynolds number and therefore in designing optimal swimmers in Newtonian fluids \cite{Optimal_sheet, sebastien_nutrient, Tam_optimal, optimal_yeomans}.

In many instances, microswimmers swim in fluids which are not Newtonian and show complex rheological properties \cite{arratia_review}. Among others, one example is of a mammalian sperm in the female reproductive tract \cite{lisa_fauci} where cervical mucus displays viscoelasticity and shear-thinning viscosity \cite{Lai_mucus}. Consequently, several model swimmers studied in Newtonian fluids have also been studied in non-Newtonian fluids for a comparison of their swimming dynamics \cite{propulsion_eric, powers_gels, gaffney2013, montenegro, jfmCharu,hewitt_balmforth_2017,datt17}.  The change in the swimmer's dynamics -- whether a change in its propulsion velocity for a fixed swimming gait or a change in the gait itself for either a fixed actuation force or fixed energy consumption-- is found to be swimmer dependent \cite{elfring2015theory} and in general we see that it is fraught with peril to generalize results obtained for one swimmer to others \cite{jfmCharu,Gaurav}. Perhaps more interestingly, and closer to the present work, are strategies that do not lead to swimming in Newtonian fluids but can be useful in complex fluids. \citet{deborah_number} first showed this for a squirmer with a surface velocity distribution that does not lead to any net motion over one cycle in a Newtonian fluid, but does so in a viscoelastic fluid. \citet{keim12} then demonstrated experimentally this elasticity enabled locomotion for a rigid assembly of two connected spheres undergoing rotational oscillations about an axis perpendicular to their mutual axis of symmetry. \citet{Bohme_2015} observed the same for axisymmetric swimmers performing reciprocal torsional oscillations. \citet{lailai_snowman} modelled a snowman swimmer, which has two unequal spheres that rotate about their common axis, that can swim only in complex fluids. Indeed it is known that the scallop theorem does not hold in complex fluids \cite{lauga2011}; fluid inertia, nearby surfaces, elasticity of the swimmer body, or interaction with other swimmers are some other reasons why a reciprocal gait for a swimmer may lead to net motion \cite{lauga2011}. In truth, the motivation for this work came from the interesting experimental and computational works of \citet{klotsa_2015} and \citet{jones2018transition} who show that an assembly of two rigid collinear spheres with a single degree of freedom can swim in the presence of inertia, and can in fact also reverse its direction at higher Reynolds number.  \citet{Felderhof2016} then theoretically studied the effect of inertia on the motion of such collinear swimmers.

In this work, we consider two different two-sphere `swimmers'. The first is simply an assembly of two spheres connected as a rigid body that is then oscillated by some external force that is aligned along the axis of symmetry of the two spheres. Strictly speaking, this is not a swimmer because the motion of the body arises as a consequence of the external force; however, we will see that by imposing a sinusoidally varying force (with zero mean value) we can achieve rectified `swimming' motion in a complex fluid. This is similar to the two-sphere system developed by \citet{lailai_snowman} that achieved net motion under an imposed torque exerted by an external (magnetic) field, although imposing an oscillatory force is perhaps easier to accomplish experimentally. The second swimmer is a two-sphere assembly where the swimming gait is prescribed as the sinusoidal variation of the distance between the two spheres with no imposed external force. This is similar to the Najafi-Golestanian swimmer \cite{najafi2004} except that here instead of three spheres we have only two and a single degree of freedom.

We emphasize that neither of these swimmers can achieve any net motion over a complete cycle in a Newtonian fluid at zero Reynolds number, irrespective of the radii of the spheres. This is due to the reciprocal forcing of the first swimmer and the reciprocal prescribed swimming gait of the second \cite{purcell1977}. In contrast, we will show that in a viscoelastic fluid, both swimmers move in the direction of the smaller sphere when the spheres are of unequal radius and nowhere if the spheres are identical. This motion is a nonlinear viscoelastic response elicited from the deformation of the microstructure of the fluid and is therefore absent in Newtonian fluids. In light of this, a two-sphere assembly in a viscoelastic fluid may also be used as a micro-rheometer as previously demonstrated in the works of \citet{Aditya_micro} and \citet{lailai_snowman}, but an assembly of two rigidly connected spheres oscillating in a fluid is perhaps the simplest such example of a nonlinear micro-rheometer. Here we use the method of perturbation expansion to study the two-sphere swimmers in an Oldroyd-B fluid which for small extension rates is a reasonable approximation of polymeric fluids \cite{propulsion_eric}. To conclude this work we compare our results with another two-sphere swimmer wherein the spheres themselves deform elastically in a Newtonian fluid -- a comparison of two-sphere swimmers in the presence of elasticity, either of the fluid or the solid.

\section{Swimmer in a viscoelastic fluid}
\label{fluid_elasticity}

\subsection{Two-sphere swimmers}
\begin{figure}
\includegraphics[width = 0.3\textwidth]{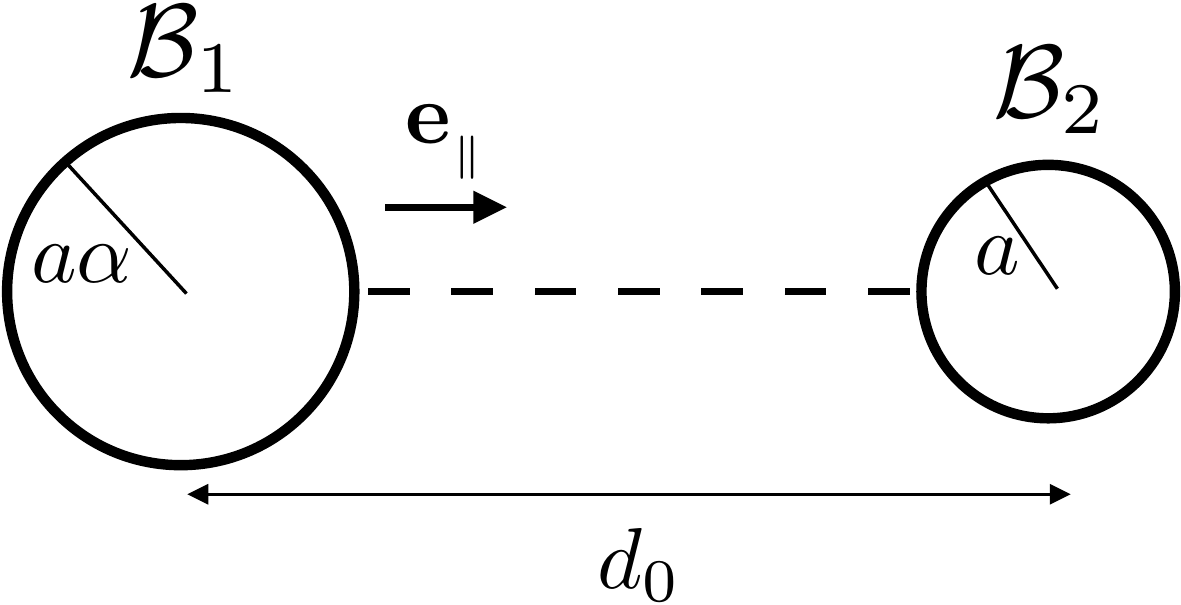}
\caption{Schematic of the two-sphere swimmer. The spheres labeled $\mathcal{B}_1$ and $\mathcal{B}_2$ of radii $a\alpha$ and $\alpha$, respectively with $\alpha>1$. The spheres are (on average) a distance $d_0$ apart. $e_{\para} $ is the unit vector pointing from $\mathcal{B}_1$ to $\mathcal{B}_2$.}
\label{figure1}
\end{figure}

In order to describe the motion of a swimming object, we decompose the contributions of the velocity of the body, 
\begin{align}
\bv(\bx\in\partial\fB) = \bU+\bOmega\times\br+\bv^S,
\end{align}
where $\bU$ and $\bOmega$ are the rigid-body translation and rotation, and the swimming gait is denoted $\bv^S$. Here the body $\fB$, with boundary $\partial\fB$, is composed of two spheres of radius $a$ and $\alpha a$, labeled $\fB_2$ and $\fB_1$ respectively ($\fB = \fB_1\cup\fB_2$). Without lack of generality we assume $\alpha \ge 1$. The distance between the two spheres is $d$, which is directed along the $\be_\para$ (from large to small sphere) as shown in figure \ref{figure1}.

\begin{figure}
\includegraphics[width = 0.5\textwidth]{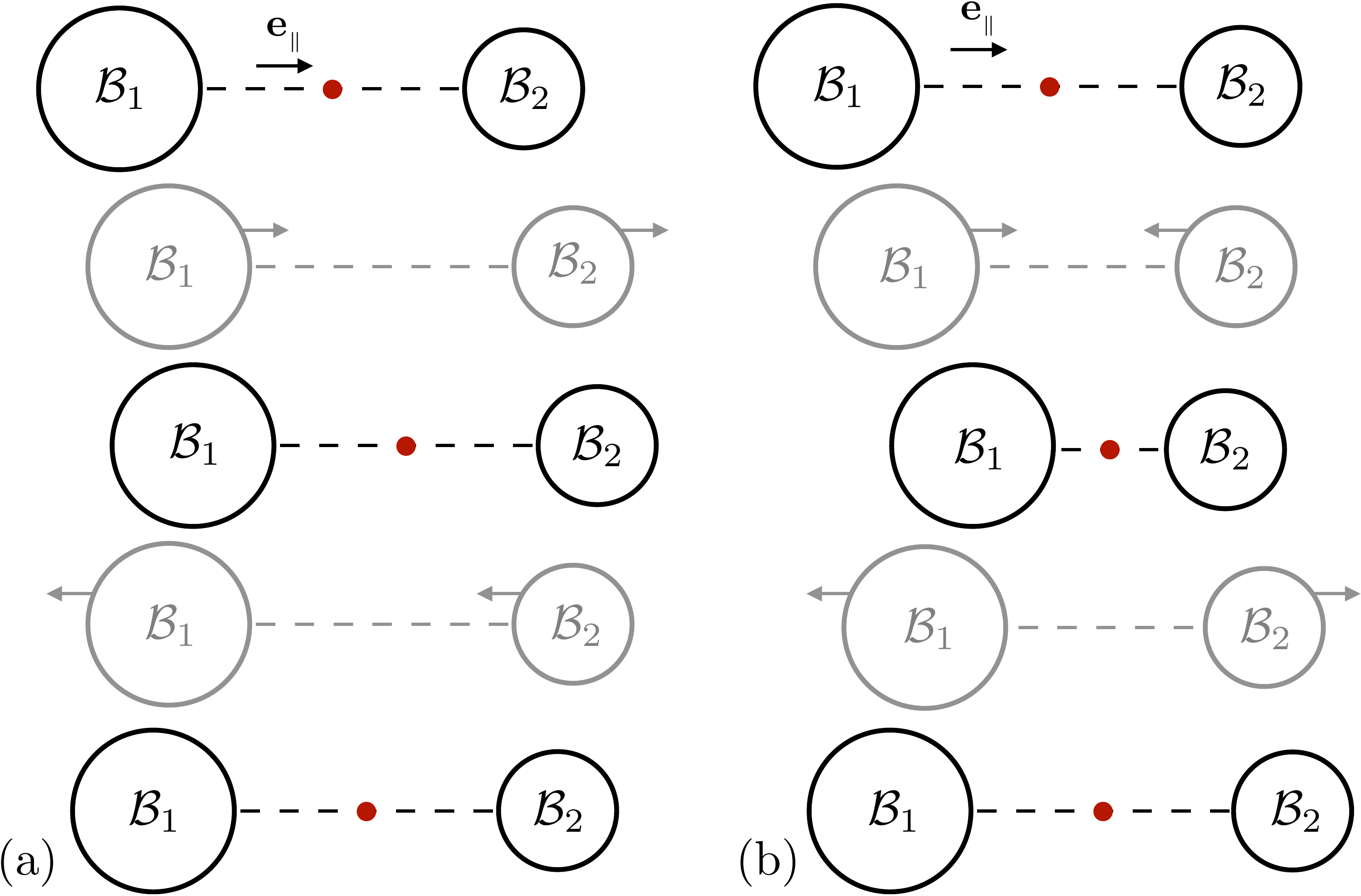}
\caption{Schematic showing one complete cycle for the two swimmers: (a) The \textit{in-phase} swimmer maintains the distance between the spheres as it moves forward. (b) In the \textit{anti-phase} swimmer, the spheres converge and diverge.  The steps in grey show the transition from one half cycle to the next. The red dot marks the position of the swimmer.}
\label{figure2}
\end{figure}

When the two spheres are connected as a rigid body the distance between the two sphere centers is a fixed constant $d=d_0$, there is no swimming gait $\bv^S=\bm{0}$, but we apply an oscillatory external force
\begin{align}
\bF_{ext} = F \cos(\omega t)\be_\para.
\label{fc}
\end{align}
This may be imposed by applying an oscillating external magnetic field if the spheres are magnetic, or if the spheres are not density matched with the fluid, simply by oscillating the medium (although in that case there would be a mean force we well). We will refer to this as an \textit{in-phase} swimmer because the two spheres move in unison (see figure \ref{figure2}a).

In contrast to the first swimmer, the distance $d$ between the spheres of second swimmer varies sinusoidally according to
\begin{align}
d = d_0 + 2\delta \sin(\omega t),
\end{align}
as equal and opposite velocities are imposed on the two spheres
\begin{align}
\bv^S(\bx\in\partial \fB_1) &= \delta \omega \cos(\omega t) \be_\para,\label{bc1a}\\
\bv^S(\bx\in\partial \fB_2) &= -\delta \omega \cos(\omega t) \be_\para\label{bc1b}.
\end{align}
Here $d_0$ is the average distance, $\delta$ is the amplitude of oscillation and $\omega$ is the frequency.  We refer to this swimmer as the \textit{anti-phase} swimmer (see figure \ref{figure2}b).

For the sake of comparison between the two swimmers, we set the magnitude of the force $F$ in \eqref{fc} such that to leading order magnitude of the velocity of the induced oscillations would be $\delta \omega$ (see appendix for further details).

\subsection{Theory for swimming in complex fluids}
The motion $\tU$ of an arbitrary swimmer (or active particle) in a non-Newtonian fluid, with deviatoric stress
\begin{align}
\btau = \eta\bgammad + \btau_{NN},
\end{align}
where $\btau_{NN}$ is the additional non-Newtonian stress, at zero Reynolds number is given by
\begin{align}\label{reciprocalmain}
\tU=\frac{\etah}{\eta}\tRh_{\tF\tU}^{-1}\cdot\lb[\tF_{ext}+\tF_T+\tF_{NN}\rb].
\end{align}
where $\tU=[\bm{U} \ \bm{\Omega}]$ is six-dimensional vector comprising rigid-body translational and rotational velocities respectively (we use bold sans serif fonts for six-dimensional vectors and tensors and bold serif for three dimensional ones) \cite{elfring2017force, DATT2018}.  The six-dimensional vector $\tF_{ext}=[\bF_{ext} \ \bL_{ext}]$ contains any external force and torque acting on the swimmer. The term
\begin{align}
\tF_T = \frac{\eta}{\etah}\int_{\partial\fB}(\bv^S-\bv^\infty)\cdot(\bn\cdot\tTh_\tU)\d S,
\label{thrust}
\end{align}
is a Newtonian `thrust' due to any surface deformation $\bv^S$ of the swimmer in a background flow $\bv^\infty$. Here we consider an otherwise quiescent fluid so that $\bv^\infty=\bzero$. The non-Newtonian contribution
\begin{align}\label{forcenn}
\tF_{NN} = -\int_{\fV}\btau_{NN}:\tEh_{\tU}\d V,
\end{align}
represents the extra force/torque on each particle due to a non-Newtonian deviatoric stress $\btau_{NN}$ in the fluid volume $\fV$ in which the particles are immersed.

These formulae rely on operators from a resistance/mobility problem in a Newtonian fluid (with viscosity $\hat{\eta}$)
\begin{align}
\bgammadh &= 2\tEh_{\tU}\cdot\tUh,\label{linearE}\\
\bsigmah &= \tTh_{\tU}\cdot\tUh,\label{linearT}\\
\tFh &= -\tRh_{\tF\tU}\cdot\tUh\label{linearR}.
\end{align}
The tensors $\tEh_{\tU}$ and $\tTh_{\tU}$ are functions of position in space that map the rigid-body motion $\tUh$ of the swimmer to the fluid strain-rate and stress fields respectively, while the rigid-body resistance tensor
\begin{align}
\tRh_{\tF\tU} = 
\begin{bmatrix}
\bRh_{FU} & \bRh_{F\Omega}\\
\bRh_{LU} & \bRh_{L\Omega}
\end{bmatrix}.
\end{align} 
Both problems considered here are axisymmetric, with the forcing and gait aligned with the axis of symmetry. In this case the resistances are diagonal and only translational motion occurs simplifying matters substantially.

We consider here only the time-averaged or (post-transient) mean velocity of the swimmer,
\begin{align}\label{reciprocalmain}
\overline{\tU}=\frac{\etah}{\eta}\overline{\tRh_{\tF\tU}^{-1}\cdot\lb[\tF_{ext}+\tF_T+\tF_{NN}\rb]},
\end{align} 
where the overline represents a time-averaged quantity. The in-phase swimmer does not change shape therefore the resistance is constant and $\tF_T=\bzero$ because $\bv^S=\bzero$; furthermore, the prescribed force is periodic with zero mean, $\overline{\tF_{ext}}=\bzero$. In contrast, the anti-phase swimmer has no external forcing $\tF_{ext}=\bzero$, but undergoes a reciprocal shape change and so, while the resistance is not constant, we know that $\overline{\tRh_{\tF\tU}^{-1}\cdot\tF_T}=\bzero$ by the scallop theorem \cite{ishimoto12}. We see then that, for both swimmers, the net motion is only due to the non-Newtonian contribution from the rheology of the fluid medium
\begin{align}
\overline{\tU}=\frac{\etah}{\eta}\overline{\tRh_{\tF\tU}^{-1}\cdot\tF_{NN}}.
\end{align}
Furthermore, by the symmetry of the problem any net motion must be in the direction of the axis of symmetry $\be_\para$ i.e. $\overline{\tU}=\overline{U}\be_\para$ with
\begin{align}
\overline{U} &=-\frac{\etah}{\eta \Rh_{FU_\para}}\int_{\fV}\overline{\btau_{NN}}:\bEh_{U_\para}\d V,
\label{steady_vel}
\end{align} 
where $\Rh_{FU_\para}=\be_\para\cdot\bRh_{FU}\cdot\be_\para$ is the scalar resistance to translational motion of the two-sphere assembly in the direction of the axis of symmetry, whereas $\bEh_{U_\para}= \tEh_{U}\cdot\be_\para$ is a second order tensor equal to the strain-rate field due to rigid-body translation (with unit speed) in the direction $\be_\para$. $\Rh_{FU_\para}$ and $\bEh_{U_\para}$ are obtained by way of the Stimson-Jeffery solution of two spheres moving with equal velocities along their axis of symmetry in a Newtonian fluid \cite{stimson1926motion}. Finally, we note that although the geometry of the anti-phase swimmer is not constant, we solve the problem asymptotically for small deformations about a mean geometry such that $\Rh_{FU_\para}$, $\bEh_{U_\para}$ and boundary of the volume integral in \eqref{steady_vel} are constant, which allows us to pass the time-average operator onto the non-Newtonian stress alone \cite{Gaurav, Lauga_theorem}.

\subsection{Constitutive equation}
We are interested here in the effects of nonlinear viscoelasticity that enable the net motion of the swimmers. Until this point we have only assumed that the stress in the fluid may be separated into a Newtonian and non-Newtonian contribution. The deviatoric stress $\btau_{NN}$ in a viscoelastic fluid typically follows a nonlinear evolution equation. For simplicity, we use the Oldroyd-B constitutive equation \cite{bird1987dynamics} but other constitutive relationships can be easily used within this formalism. Oldroyd-B is a single relaxation time viscoelastic (Boger fluid) fluid that is governed by
\begin{align}
\buildrel \nabla \over \btau_{NN}= \frac{\eta_{NN}}{\lambda}\bgammad-\frac{1}{\lambda}\btau_{NN},
\end{align}
where $\lambda$ is the relaxation time of the fluid and $\eta_{NN}$ is an additional viscosity due to the (polymeric) microstructure. The upper convected derivative is defined $\buildrel\nabla\over {\bA} = \partial \bA/\partial t + \bv \cdot \bnabla \bA - \left(  \left(\bnabla \bv\right)^T \cdot \bA + \bA \cdot \bnabla \bv \right)$ where $\bv$ is the fluid velocity field.

The problems we consider here are periodic (with period $\tau = 2\pi/\omega$) and, neglecting any transient evolution from an initial condition, we may simplify matters by assuming that all functions may be written as Fourier series, for example, the velocity field $\bv = \sum_p \bv^{(p)}e^{pi\omega t}$. Following this for the stress, we have \cite{Gaurav}
\begin{align}
\btau^{(p)}_{NN}= (\eta^*(p)-\eta)\bgammad^{(p)}+\bN^{(p)}
\label{tauNN}
\end{align} 
where the tensor $\bN^{(p)}$ represents the contribution of the nonlinear terms to each mode and the complex viscosity
\begin{align}
\eta^*(p) = \frac{1+pi\De \beta}{1+pi\De}\eta_0.
\label{complexviscosity}
\end{align}
The Deborah number, $\De = \lambda \omega$, characterizes the relative rate of actuation of the spheres to the relaxation of the fluid. The viscosity ratio $\beta = \eta / \eta_0$ is the relative viscosity of the Newtonian part of the fluid (solvent) where $\eta_0=\eta+\eta_{NN}$ represents the (total) zero-shear-rate viscosity of the fluid. In particular, by substituting \eqref{tauNN} into \eqref{steady_vel} one may show that
\begin{align}
\overline{U} &=-\frac{\etah}{\eta_0 \Rh_{FU_\para}}\int_{\fV}\overline{\bN}:\bEh_{U_\para}\d V,
\end{align} 
where $\overline{\bN}=\bN^{(0)}$, and we see that linear viscoelasticity does not lead to net motion of these swimmers because by definition $\bN^{\left( p\right)}=\bzero$ for linearly viscoelastic fluids (see the appendix for further details).

\subsection{Small amplitude expansion}
We assume that the oscillation amplitudes are much smaller than all other length scales, $\delta \ll a, d_0$, and define dimensionless quantities $\epsilon = \delta/a \ll 1$ and $\Delta = d_0/a$. In addition we define a dimensionless clearance between the spheres, $\Delta_c = \Delta -(1+\alpha)$.  We solve for the flow by employing a regular perturbation expansion in small deformations $\epsilon$ to all flow quantities
\begin{align}
\left\{ \bv, \btau, p, \ldots \right\} = \epsilon \left\{ \bv_1, \btau_1, p_1, \ldots \right\} + \epsilon^2 \left\{ \bv_2, \btau_2, p_2, \ldots \right\} + \ldots. 
\end{align}
The swimming speed is then given by
\begin{align}
\overline{U} &=-\epsilon^2\frac{\etah}{\eta_0 \Rh_{FU_\para}}\int_{\fV}\overline{\bN_2}:\bEh_{U_\para}\d V+\fO(\epsilon^4),
\label{vel_epsilonsquare}
\end{align} 
Because the tensor $\bN$ represents the nonlinear terms in the viscoelastic constitutive equation there are no terms linear in $\epsilon$. The quadratic term depends only on the leading order flow field, $\bN_2[\bv_1,\btau_1]$, which is a solution to a linearly viscoelastic flow that has exactly the same flow field as a Newtonian flow with equivalent prescribed velocity boundary conditions.

When the spheres move together as a rigid body (the in-phase swimmer), the solution for $\bv_1$ is easily obtained using the solution for two spheres moving with equal velocities along the line joining their centers by \citet{stimson1926motion}. Similarly when the spheres approach one another (\text{anti-phase} swimmer), the solution for $\bv_1$ is available due to the work of \citet{maude1961end} for two spheres approaching each other in a Newtonian fluid  (see \cite{spielman_1970} for some corrected errors). Thus knowing the $\mathcal{O} \left( \epsilon\right)$ fields, we may evaluate the tensor $\bN_2$, which for an Oldroyd-B fluid is given by
\begin{align}
\overline{\bN_2} = - \frac{1}{2}  \Re \left\{\frac{\De \left( 1- \beta \right)}{\left(1+ \di \De\right)} \left[ \bv_1^{(-1)} \cdot \nabla \dot{\bgamma}_1^{(1)} - \left( \nabla \bv_1^{(-1)}\right)^T \cdot \dot{\bgamma}_1^{(1)} - \dot{\bgamma}_1^{(-1)}\cdot \nabla \bv_1^{(1)} \right]\right\}.
\label{steady_comp}
\end{align}
Finally we obtain the leading order motion for either swimmer by evaluating \eqref{vel_epsilonsquare} to find
\begin{align}
\overline{U}  =  \delta\omega\frac{\delta}{a}\lb(\dfrac{\De \left( 1- \beta\right)}{1+ \De^2}\rb)\mathcal{U} \be_\para, 
\label{steady_velocity_simple}
\end{align}
where the dimensionless quantity $\mathcal{U}$ is evaluated using numerical integration of an analytical expression.

\subsection{Results \& Discussion}
\begin{figure}
\includegraphics[width = 0.5\textwidth]{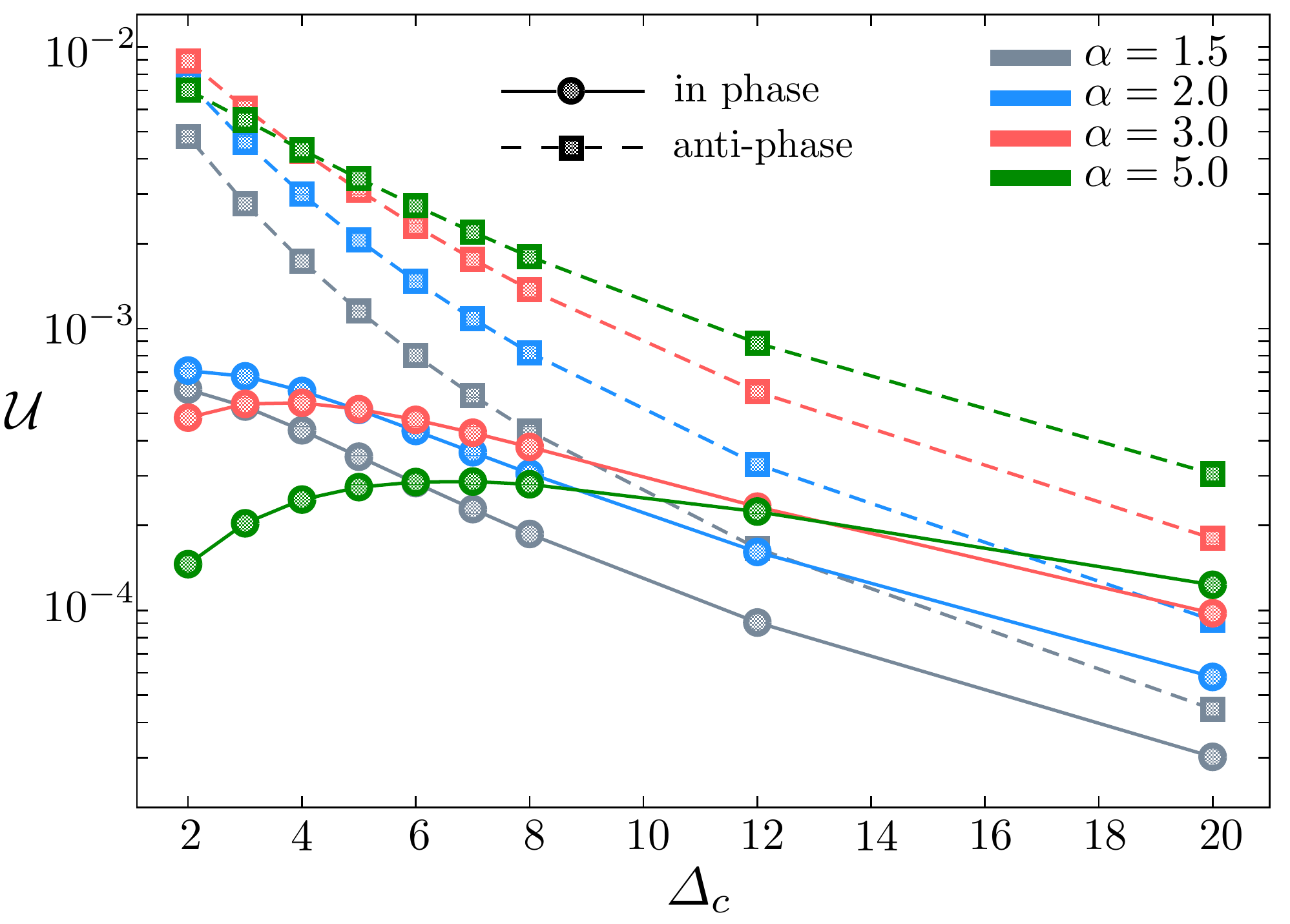}
\caption{The swimming speed coefficient $\mathcal{U}$ is plotted with variation in the clearance $\Delta_c$ between the two spheres for different size ratios $\alpha$. The square symbols (connected by dashed lines) represent the anti-phase swimmer and the circles (connected by solid lines) represent the in-phase swimmer. All quantities are dimensionless.}
\label{figure3}
\end{figure}

We find that the two-sphere assembly can swim in a viscoelastic fluid at finite Deborah numbers, provided the two spheres are of different sizes. The difference in the sphere sizes leads to the fore-aft asymmetry required for swimming. We see from \eqref{steady_velocity_simple} that the swimming speed is maximized when $\De=1$. In the limit when the actuation is much slower than the relaxation of the fluid, $\De\rightarrow 0$, or much faster, $\De\rightarrow\infty$, there is no swimming $\overline{U}=0$, indeed the term in the brackets of \eqref{steady_velocity_simple}, which governs this behavior, is simply the dimensionless elastic modulus of the fluid \cite{Gaurav}. For specific cases we report the values of $\mathcal{U}$ for the two swimmers for a few configurations in figure \ref{figure3}. Both swimmers swim with the smaller sphere as the head. At small separations the anti-phase swimmer is an order of magnitude faster; however, at large separations this difference in magnitude fades away.  

The direction of the motion of these swimmers can be largely predicted by studying a single sphere oscillating in a viscoelastic fluid. The \textit{viscoelastic steady streaming flow} that results from this motion draws fluid in towards the center of the sphere along the axis of oscillation \cite{bohme92}. Larger spheres generate stronger viscoelastic flows for a given velocity but the relationship is sublinear in radius and so one would expect that when two unequal spheres interact, because of the relative resistances, the net effect of the interacting viscoelastic streaming flows would be to push the assembly in the direction of the smaller sphere. This is essentially a `far-field' superposition argument, where there is no difference between in-phase and anti-phase oscillations, and one should take great care when applying this logic to closely interacting spheres in a nonlinear non-Newtonian fluid; however, this prediction qualitatively agrees with our exact two-body problem solutions. We also note that \citet{keim12} find that a similar two-sphere assembly undergoing rotational oscillations instead moves towards the larger sphere, but in that case the spheres are moving perpendicular to their axis of symmetry and so we expect the viscoelastic steady streaming flow to be reversed along that axis.

Examining more closely first the in-phase swimmer. A rigid body of such shape moving in a weakly viscoelastic fluid (e.g. a second-order fluid under slow flows \cite{bird1987dynamics}) will experience a net viscoelastic force pointing towards the smaller sphere and so the total drag on a body when the the larger sphere leads increases while it decreases when the smaller sphere is at the front. \citet{Leal-fore-aft} has also shown that for sedimenting slender bodies, when the trailing end is sharp and the leading edge is blunt the drag increases in a second-order fluid. In light of this, one should expect that when the two-sphere body oscillates periodically in a viscoelastic fluid, the net viscoelastic contribution to the force on the body over one cycle to point towards the smaller sphere. The speed of the swimming depends on the strength of this viscoelastic contribution and the hydrodynamic resistance to the steady translation of the body. As can be seen from figure \ref{figure3}, such a swimmer has an optimum in the swimming velocity at a certain separation for a given ratio of the sphere sizes. 

For the \text{anti-phase} swimmer, the viscoelastic force seems to depend on the strength of squeeze flow between the two spheres which increases as the separation between the spheres decreases. Combined with the low hydrodynamic resistance of the assembly when the spheres are close, swimming is monotonically faster with smaller separations (for a given size-ratio). When the spheres are far apart the strength of the squeeze flow decreases and the two types of swimmers swim with speeds of the same order. 

Clearly, a size ratio of 1 will not lead to swimming. One also expects a very large size ratio to be equally inefficient due to a decrease in the net fore-aft asymmetry over a complete cycle. This non-monotonicity with size ratio is also observed at small distances in figure \ref{figure3}, although at very large distances, when the interaction between the spheres has much decreased, higher size ratio leads to better swimming. However, this may not be the regime one would focus for optimal swimming.

We also note that the effect of viscoelasticity on the swimmers is found to be opposite to the effect of inertia as described in the analytical work of \citet{Felderhof2016}. There the two-sphere swimmer moves with the larger sphere as the head, as might be expected given that weakly inertial steady-streaming flow instead pushes fluid out from an oscillating sphere along the axis of oscillation \cite{riley66,bohme92,spelman17}. However, recent numerical work by \citet{jones2018transition} reports that the smaller sphere leads at small Reynolds number only to switch to larger-sphere leading at higher Reynolds number. We do not observe such switching of swimming direction with the Deborah number in our analysis which is valid for small oscillation amplitudes. 
 
In the next section, we study a two--sphere swimmer with elastic spheres in a Newtonian fluid and demonstrate that the direction of propulsion is the same as this two--(rigid)--sphere swimmer in viscoelastic fluid. 

\section{Swimmer with elastic spheres}
\label{elastic_spheres}
We now compare the two-sphere swimmers in a viscoelastic fluid with swimmers with elastic spheres in a Newtonian fluid. This calculation closely follows the work of \citet{Babak_2017} who studied a two-sphere swimmer with one rigid and other elastic sphere in a Newtonian fluid.   Here, similar to the previous section, we consider model swimmers that consist of two spheres of radii $a$ and $\alpha a$, but this time we relax the rigidity constraint by assuming that the spheres are isotropic, incompressible neo-Hookean solids.

To study the behavior of this system, one must first understand the deformation of a single elastic sphere in Stokes flow. Neglecting intertia, momentum balance for the elastic solid yields
\begin{align}
\label{solid1}
{\boldsymbol{\nabla}}\cdot{\boldsymbol{\sigma}}_s + {\mathbf{f}}(t)=\mathbf{0},
\end{align}
where ${\boldsymbol{\sigma}}_s$ is the stress due to elastic deformation and ${\mathbf{f}}$ is the applied body force density on the sphere. For an isotropic, incompressible neo-Hookean solid, this stress field can be expressed using the displacement vector ${\mathbf{u}}$ as \cite{ogden1984,gurtin2010}
\begin{align}
{\boldsymbol{\sigma}}_s=-{p}_s\mathbf{I}+ G \left({\boldsymbol{\mathsf{{F}}}}\cdot{\boldsymbol{\mathsf{{F}}}}^T-\mathbf{I}\right),
\end{align}
where $G$ is the shear modulus and ${\boldsymbol{\mathsf{{F}}}}=\mathbf{I}+{\boldsymbol{\nabla}}{\mathbf{u}}$ is the deformation gradient tensor. The Lagrange multiplier ${p}_s$ enforces the incompressibility of the solid through
\begin{align}
\text{det}\left({\boldsymbol{\mathsf{{F}}}}\right)=1,
\end{align}
where `det' is the determinant. The traction across the solid-fluid interface must be continuous so that
\begin{align}
\label{traction}
{\boldsymbol{\sigma}}_s\cdot\mathbf{n}={\boldsymbol{\sigma}}\cdot\mathbf{n},
\end{align}   
where $\mathbf{n}$ is the normal vector to the deformed sphere and ${\boldsymbol{\sigma}}$ is the stress field in the fluid domain which can be determined by solving the Stokes equations over the deformed boundary.

If we scale lengths with $a$, velocities with $\delta\omega$, forces with $G/a $, time with $a/\delta\omega$, stress in the solid domain with $G$ and stress in the fluid domain with $\eta\delta\omega/a$, from equation \eqref{traction} a dimensionless parameter $\varepsilon=\eta\delta\omega/aG$ then naturally arises as the ratio of viscous forces to elastic forces (recall also that $\Delta=a/d_0$). Here we focus on the case wherein the sphere is only weakly elastic; elastic forces are much larger than viscous forces and so $\varepsilon\ll 1$. Since the motion is axisymmetric, one can show that the elastic sphere reaches equilibrium with a relaxation time scale of $\tau_\text{relax}\sim \mathcal{O}(a\varepsilon/\delta\omega)$. Thus, under the assumption of $\varepsilon\ll 1$, we can assume that elastic deformations are quasi-static: the sphere deforms instantly and we then have rigid-body motion \cite{huang2011}.

Similar to the viscoelastic case, for the in-phase swimmer, for the sake of comparison we set the magnitude of the applied external force to be $F=\delta \omega R_{FU_\para}$ so that to leading order the speed of oscillation is $\delta\omega$. For the anti-phase swimmer we define the gait according to according to \eqref{bc1a}, \eqref{bc1b} but in this case the velocity is prescribed on the deformed boundaries.

We now return to our two-sphere swimmer, with both spheres being weakly elastic. In a Newtonian fluid the dynamics of the motion of the body is given by
\begin{align}
\bU = \bR_{FU}^{-1}\cdot\lb[\bF_T+\bF_{ext}\rb]
\end{align}
The thrust force may be generically decomposed into the thrust generated by each sphere $\bF_T = \bF_{T_1}+\bF_{T_2}$. Because the spheres are deforming, we will assume that the spheres are well separated, and compute the hydrodynamic thrust generated by each sphere with hydrodynamic interactions solved to leading order using a far-field approximation, $\Delta\ll 1$.

For individual spheres \eqref{thrust} reduces to Fax\'en's first law for each sphere
\begin{align}
\bF_{T_1}&=-\bR_1\cdot\left(\bv_1^S -\mathcal{F}_1\left[\bv_2^\infty\right]\right),\\
\bF_{T_2}&=-\bR_2\cdot\left(\bv_2^S -\mathcal{F}_2\left[\bv_1^\infty\right]\right),
\end{align}
where $\bR_1$ and $\bR_2$ are the resistance tensors for each sphere and, $\mathcal{F}_1$ and $\mathcal{F}_2$ are the respective Fax\'en operators. Here, $\bv_1^\infty$ is the background flow field induced by sphere $\fB_1$, and vice versa for $\bv_2^\infty$. Recalling that spheres are only weakly elastic (since $\varepsilon\ll 1$), the spheres only slightly deviate from their spherical shape so that the hydrodynamic resistance and Fax\'en's laws are unchanged from an undeformed sphere to leading order \cite{brenner1964,kim1985}. The net thrust generated by the swimmer at the leading order is thereby
\begin{align}
\bF_T=6\pi\eta a\left(-\alpha\bv_1^S +\alpha\bv_{2,1}^\infty- \bv_2^S +  \bv_{1,2}^\infty\right),
\end{align}
where $\bv_{2,1}^\infty$ indicates the background flow from spheres 2 evaluated at the center of sphere 1 (and vice versa). For the externally forced swimmer, the gait is zero $\bv_1^S=\bv_2^S=\bzero$. For the anti-phase swimmer, the imposed gait is periodic and given that we are interested in only the mean motion, averaging over a period $\tau=2\pi/\omega$, in either case, leads to
\begin{align}
\overline{\bF_T}=6\pi\eta a\left(\alpha\overline{\bv_{2,1}^\infty}+\overline{\bv_{1,2}^\infty}\right).
\end{align}
We see clearly, in this far field result, that the thrust is dictated purely by the \textit{elastic steady streaming flow} generated by each sphere acting on the other. 

Now by solving equations \eqref{solid1} to \eqref{traction} asymptotically, one can determine the flow field around an oscillating elastic sphere, then averaging to obtain the steady streaming flows $\overline{\bv_1^\infty}$ and $\overline{\bv_2^\infty}$ (see \cite{Babak_2017} for technical details). By prescribing an external force magnitude set by $F=\delta\omega R_{FU\para}$ the magnitude of the deformation and thus the magnitude of the steady streaming flows is equal for both swimmers. We note in particular that the the elastic steady streaming flow of each sphere draws fluid inward along the axis of symmetry in much the same way as the viscoelastic steady streaming flow. Here we find that $\bv_{2,1}^\infty\cdot\be_\para\propto \delta\omega \Delta^2\epsilon^3$ and $\bv_{1,2}^\infty\cdot\be_\para\propto -\delta\omega \Delta^2\epsilon^3/\alpha$. The net thrust is then
\begin{align} 
\overline{\bF_T}&=\frac{74979}{34048}\pi\eta d_0 \delta\omega\alpha\left(1-\frac{1}{\alpha^2}\right)\Delta^3\epsilon^3\be_\para.
\end{align}
Both oscillating elastic spheres generate steady streaming flows but the magnitude of each flow is inversely proportional to the radius while the resistance of each sphere is linearly proportional to the radius and so the net thrust force is in the direction of the smaller sphere ($\alpha\ge1$).

With a hydrodynamic resistance of $R_{FU_\para}=6\pi \eta a (1+\alpha)$, and using the fact that the average external force is zero,
\begin{align}
\overline{\bF_{ext}}=\bzero,
\end{align}
(in the case of the anti-phase swimmer the prescribed force itself is zero), we obtain the time-averaged velocity
\begin{align} 
\overline{\bU}&=\frac{24993}{68096}\pi\eta \delta\omega\left(1-\frac{1}{\alpha}\right)\Delta^2\epsilon^3\be_\para.
\end{align}
The swimming motion is always in the direction of the smaller sphere, similar to the rigid swimmer in the viscoelastic fluid (the swimmer swims with the smaller sphere as the head). Furthermore, since we solved this problem assuming the spheres are well separated using far-field approximations of the flow, the speed of the swimmer is ultimately independent of whether the spheres oscillate in phase or anti-phase.  

\section{Conclusion}
\label{conclu}

We studied the effects of elasticity on the motion of two-sphere swimmers where the two spheres oscillate in-line. When the two spheres are rigid and the fluid viscoelastic, we find that the swimmers swim with the smaller-sphere as the head. However, the swimming speed is dependent on the type of swimmer: anti-phase swimmers, in general, swim faster than the in-phase swimmers. We also find that when the spheres themselves are elastic and the fluid Newtonian, the swimmer again moves in the direction of the smaller sphere.

We note that the effects of elasticity on the swimmer are found to be opposite of the effect of inertia described in the theoretical work of \citet{Felderhof2016} who showed the two-sphere swimmer moves with the larger sphere as the head, but we do not observe a reversal swimming direction as a function of the Deborah number, analogous to what is observed upon increasing Reynolds number in the numerical work of \citet{jones2018transition}.

\section*{Acknowledgement}
Funding from the Natural Sciences and Engineering Research Council of Canada (NSERC) is gratefully acknowledged.
\appendix*
\section{Linear viscoelasticity}
Equation \eqref{reciprocalmain} delineates a relationship between forces and velocities, now substituting in \eqref{tauNN} we obtain for each Fourier mode
\begin{align}
\frac{\etah}{\eta^*(p)}\tRh_{\tF\tU}\cdot\tU^{(p)}=\tF_{ext}^{(p)}+\frac{\eta^*(p)}{\etah}\int_{\partial\fB}\bv^{S(p)}\cdot(\bn\cdot\tTh_\tU)\d S-\int_{\fV}\bN^{(p)}:\tEh_{\tU}\d V.
\end{align}

For a rigid-body motion under periodic external forcing $\bv^S=\bzero$. Assuming that the magnitude of the forcing is small so that nonlinear viscoelastic terms are negligible to leading order, we obtain (complex) linear viscoelastic relationship between force and velocity for each mode
\begin{align}
\tR_{\tF\tU}^{*(p)}\cdot\tU^{(p)}=\tF_{ext}^{(p)}
\end{align}
where the complex resistance $\tR_{\tF\tU}^*=\frac{\eta^*}{\etah}\tRh_{\tF\tU}$.

In our problem the there is only a single force mode $2F^{(1)}=F$ (else zero). Setting the magnitude of the velocity to be $|U|=\delta\omega$ then leads to a force with magnitude $F=\delta\omega |\eta^*(1)|\Rh_{FU_\para}/\etah$. Using the complex viscosity of Oldroyd-B we obtain that taking $F=\delta\omega\eta_0 \frac{1+\beta De^2}{1+De^2}\Rh_{FU_\para}/\etah$ leads to a velocity $\bU=\delta\omega\cos(\omega t+\phi)\be_\para$ to leading order.

\bibliography{reference}

\end{document}